\documentclass[letterpaper,twocolumn,prl,aps,superscriptaddress,amsmath,amssymb,floatfix]{revtex4-1}
\usepackage{mathptmx}
\usepackage[latin9]{luainputenc}
\usepackage{color}
\usepackage{amsmath}
\usepackage{amssymb}
\usepackage{graphicx}
\usepackage{esint}
\usepackage[unicode=true,
 bookmarks=true,bookmarksnumbered=false,bookmarksopen=false,
 breaklinks=false,pdfborder={0 0 1},backref=false,colorlinks=true]
 {hyperref}
\hypersetup{
 linkcolor=magenta,urlcolor=blue,citecolor=blue,pdfstartview={FitH},hyperfootnotes=false}

\makeatletter

\usepackage{textcomp}
\usepackage{epstopdf}

\usepackage{amsfonts}

\pdfpageheight\paperheight
\pdfpagewidth\paperwidth

\@ifundefined{textcolor}{}{%
 \definecolor{BLACK}{gray}{0}
 \definecolor{WHITE}{gray}{1}
 \definecolor{RED}{rgb}{1,0,0}
 \definecolor{GREEN}{rgb}{0,1,0}
 \definecolor{BLUE}{rgb}{0,0,1}
 \definecolor{CYAN}{cmyk}{1,0,0,0}
 \definecolor{MAGENTA}{cmyk}{0,1,0,0}
 \definecolor{YELLOW}{cmyk}{0,0,1,0}
}

\usepackage{xcolor}\usepackage{soul}
\definecolor{blue}{rgb}{0,0,1}
\definecolor{red}{rgb}{1,0,0}
\definecolor{green}{rgb}{0,1,0}

\usepackage{soul}

\makeatother

\begin{document}
\title{Massive particle acceleration on a photonic chip via spatial-temporal modulation}
\author{Mai~Zhang}
\thanks{These two authors contributed equally to this work.}
\affiliation{CAS Key Laboratory of Quantum Information, University of Science and Technology of China, Hefei, Anhui 230026, People's Republic of China}
\affiliation{CAS Center For Excellence in Quantum Information and Quantum Physics, University of Science and Technology of China, Hefei, Anhui 230026, China}
\author{Xie-hang~Yu}
\thanks{These two authors contributed equally to this work.}
\affiliation{CAS Key Laboratory of Quantum Information, University of Science and
Technology of China, Hefei, Anhui 230026, People's Republic of China}
\affiliation{CAS Center For Excellence in Quantum Information and Quantum Physics,
University of Science and Technology of China, Hefei, Anhui 230026,
China}
\author{Xin-Biao Xu}
\affiliation{CAS Key Laboratory of Quantum Information, University of Science and
Technology of China, Hefei, Anhui 230026, People's Republic of China}
\affiliation{CAS Center For Excellence in Quantum Information and Quantum Physics,
University of Science and Technology of China, Hefei, Anhui 230026,
China}

\author{Guang-Can Guo}
\affiliation{CAS Key Laboratory of Quantum Information, University of Science and
Technology of China, Hefei, Anhui 230026, People's Republic of China}
\affiliation{CAS Center For Excellence in Quantum Information and Quantum Physics,
University of Science and Technology of China, Hefei, Anhui 230026,
China}
\author{Chang-Ling Zou}
\email{clzou321@ustc.edu.cn}
\affiliation{CAS Key Laboratory of Quantum Information, University of Science and
Technology of China, Hefei, Anhui 230026, People's Republic of China}
\affiliation{CAS Center For Excellence in Quantum Information and Quantum Physics,
University of Science and Technology of China, Hefei, Anhui 230026,
China}
\date{\today}
\begin{abstract}
Recently, the spectral manipulation of single photons has been achieved through spatial-temporal modulation of the optical refractive index. Here, we generalize this mechanism to massive particles, i.e. realizing the acceleration or deceleration of particles through the spatial-temporal modulation of potential induced by lasers. On a photonic integrated chip, we propose a MeV-magnitude acceleration by distributed modulation units driven by lasers. The mechanism could also be applied to atom trapping, which promises a millimeter-scale decelerator to trap atoms. The spatial-temporal modulation approach is universal and could be generalized to other systems, which may play a significant role in hybrid photonic chip and microscale particle manipulation.
\end{abstract}
\maketitle

The photonic chip offers a united platform to investigate and utilize the light-matter interactions. This compact, stable, and scalable platform allows the integration of thousands of functional photonic devices on a single chip~\citep{Atabaki2018,Meng2021,Qiang2018}. More importantly, the light-matter interaction could be greatly enhanced on the chip, due to the strong confinement of optical fields in photonic micro-structures with a nanoscale cross-section~\citep{Jahani2016,Hu2021,McNeur2018,Sapra2020,Shiloh2021a}, and also the resonant-enhancement~\citep{Kfir2020,Henke2021,Kozak2021} in microcavities. Beneficial from these advantages of the photonic chip, recently, the manipulation of the optical photon has been realized in tens of distributed optomechanical waveguides~\citep{Fan2019}. It was demonstrated that when certain spatial-temporal modulation is applied to optical photons, their spectral features could be manipulated with very high efficiency.

It is anticipated that the mechanism could be extend to massive particles by applying spatial-temporal modulation through optical fields. For instance, the optical dipole force could be applied to trapping and transporting nanoparticles or atoms~\citep{Ashkin1970,Leibfried2003,Renaut2013,Ashkin1986,Wieman1999,Kaufman2012,Beugnon2007}. Recently, the optical fields in photonic microstructures have been applied to manipulate free electrons~\citep{Breuer2013,Peralta2013,England2014a}, which provide new tools to investigate and control both photons and electrons~\citep{Feist2015,Dahan2020,Dahan2021,Wang2020,Adiv2021,Zhang2021,Baranes2022}. For the the manipulation of photons, the interaction could accumulation with the propagation of photons when phase-matching condition of nonlinear optics effect is satisfied for a given device and working wavelength~\citep{Boyd2008}. However, for the manipulation of massive particles, the coupling between the massive particle and optical fields varies with the propagation of particle as its velocity changed~\citep{Plettner2006}. Therefore, the investigation the corresponding phase-matching condition is crucial for manipulating massive particles by distributed spatial-temporal modulations.

\begin{figure}
\includegraphics[width=1\columnwidth]{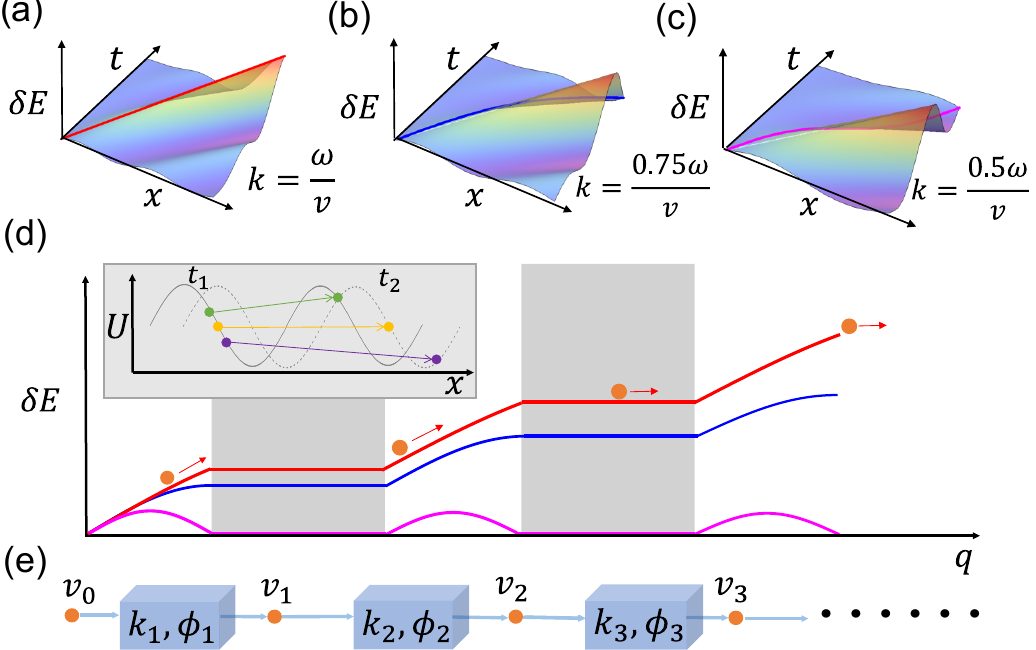}
\caption{(a)-(c) The self-energy change of the particle traveling in a typical spatial-temporal potential $U\left(x,t\right)=A\cos(\omega t-kx+\phi)$ for different wavevector $k=\omega/v,\,0.75\omega/v,\,0.5\omega/v$, respectively, with $\omega$ is angular frequency of the harmonic oscillation, $v$ is the particle velocity, $t$ is time, and $x$ is the position, and $\phi$ is the phase. The lines indicate the trajectories of the particle $x=vt$. (d) The self-energy change $\delta E$ of the moving particle in cascaded acceleration units. Here, the red, blue and magenta lines are corresponding to the potential functions in (a)-(c), with $q$ is the position of the particle on its world-line. Inset: The change of potential $U$ for particle velocity $v$ equal, slower, and faster than the group velocity of the spatial-temporal modulation of the potential. (e) The general setup for massive particle manipulation via cascaded spatial-temporal modulation units, with $k_{i}$ and $\phi_{i}$ represent the wavevector and phase parameters of $i$-th unit, $v_{i}$ is the velocity the particle exiting from $i$-th unit.}
\label{Fig1}
\end{figure}

In this Letter, we generalize the theory of spatial-temporal modulation to the interaction among massive particles and space-time varying potential. A spatial-temporal matching condition is derived for massive particles and applied to efficiently accelerate electrons by distributed unites on a photonic chip. It is demonstrated a portion of $10\%$ electrons in an initially Gaussian distributed ensemble can be accelerated by 200~keV within a length less than $6\,\mathrm{mm}$. First of all, we
explain the general theory of spatial-temporal modulation by the
experimentally verified optical frequency manipulation on a photonic
chip~\citep{Fan2016}, where photons traveling in a waveguide
while there are distributed mechanical vibrations changing the effective refractive index of the waveguide ($n_{\mathrm{eff}}$). By treating the mechanical motion  as a slowly varying external potential when photons passing through the waveguide, the photon energy is adiabatically shifted as~\citep{sm}
\begin{equation}
\delta E\approx\hbar v_{\mathrm{g}}k_{0}\int d\zeta\frac{\partial}{\partial x}\left[\Delta n_{\mathrm{eff}}\left(x,t\right)\right].\label{eq:3}
\end{equation}
Here, $\hbar$ is the reduced Plank constant, $v_{\mathrm{g}}$ is the light group velocity in the waveguide, $k_{0}$ is the vacuum photon wavevector, $\Delta n_{\mathrm{eff}}\left(x,t\right)$ is the mechanically-induced spatial-temporal modulation of $n_{\mathrm{eff}}$. The manipulation of photon self-energy can be written in a general form as
\begin{equation}
\delta E=-\int\frac{\partial U(x,t)}{\partial x}d\zeta,\label{worldline}
\end{equation}
which equals to the integral of the spatial-temporal potential $U\left(x,t\right)=-v_{\mathrm{g}}\hbar k_{0}\Delta n_{\mathrm{eff}}\left(x,t\right)$ along its trajectory $\zeta$. This expression works for both massless and massive particle to describe the change of travelling particle's self-energy by a spatial-temporal potential. For massless particles, such as photons, the self-energy manipulation corresponds to the frequency shift, while for massive particles, such as electrons and atoms, the self-energy manipulation corresponds to the acceleration or deceleration.

\begin{figure*}
\includegraphics[width=1\textwidth]{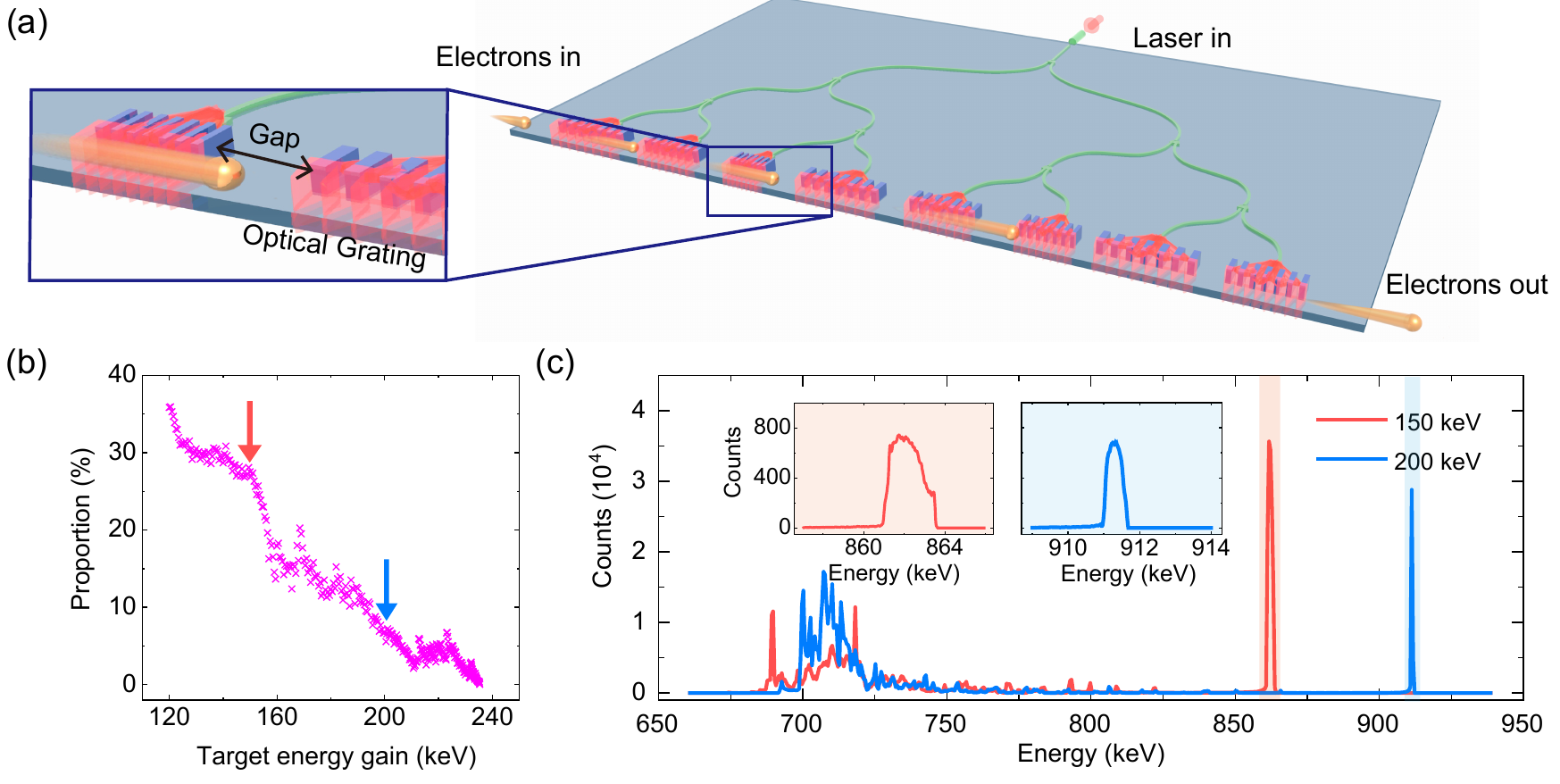}
\caption{(a) Sketch of the experimental setup for cascaded electron accelerator on a photonic chip. (b) The results of the proportion that the electron could gain energy that exceeds the target (horizontal axis), with the parameters of all units are optimized numerically and the initial energy distribution of the electron ensembles is set as a Gaussian function with a center at $711\,$keV with a standard deviation of $0.14\,\mathrm{keV}$ and the initial phase is random. The electrical field strength is $5\times10^{8}\,\mathrm{V/m}$, the unit length is $5.21\,\mathrm{\mu m}$, and the initial wavevectors are set as $k=5.753\,\mathrm{\mu m}^{-1}$. (c) The final distribution of $500\mathrm{k}$ input electrons for different target energy gain, here the energy interval is $0.5\,\mathrm{keV}$. The Red line shows the result when target energy gain is $150\,\mathrm{keV}$, and the blue line shows the $200\,\mathrm{keV}$situation.  Inset: distribution around two peaks, here the energy interval is $10\,\mathrm{eV}$. }
\label{Fig2}
\end{figure*}

In a typical spatial-temporal dependent potential $U\left(x,t\right)=A\cos(\omega t-kx+\phi_{0})$,  with $A$, $\omega$, $k$ and $\phi_{0}$ denote the amplitude, angular frequency, wave vector and initial phase of the modulation, the particle interacting with the harmonically modulating potential is shown in the inset of Fig.~\ref{Fig1}(d). For convenience, the potential $U(x,t)$ can be written as a function of the position of the world-line $q\left(t\right)=x_{0}+v_{0}(t-t_{0})$, with $v_{0}$ is the initial velocity and $t_{0}$ $\left(x_{0}\right)$ is the initial time (location). In a single modulation unit of length $L$ with uniform $A$ and $k$, the self-energy change could be approximated to the first order, i.e. the change of the particle velocity within the unit is negligible, as
\begin{equation}
\delta E\left(L\right)=-\frac{2Akv_{0}}{\omega-kv_{0}}\sin[\frac{\omega L}{2v_{0}}-\frac{kL}{2}+\phi_{0}]\sin[\frac{\omega L}{2v_{0}}-\frac{kL}{2}],\label{eq:STMC}
\end{equation}

Shown in Figs.~\ref{Fig1}(a)-(c) are the illustrations of typical self-energy change of massive particles under different conditions for $k=\omega/v,\,0.75\omega/v,\,0.5\omega/v$. It is found that the optimal energy gain is achieved for $k=\omega/v$, while the other conditions give energy gain with a certain $L$ but reduce by further increasing the $L$. From Eq.~(\ref{eq:STMC}), it is obvious that when $\phi_{0}=3\pi/2$ and $\omega-kv_{0}=0$, i.e. the spatial-temporal matching (STM) condition between the particle and the potential, is fulfilled, the energy gain could be achieved as $\delta E=AkL$. When deviating from this STM condition, $\delta E$ is a trigonometric function that oscillates with $L$ and is bounded to $\delta E\leq\frac{2Akv_{0}}{\omega-kv_{0}}\sin^{2}\frac{\phi_{0}}{2}$, and the bound is achieved when $L=\frac{\phi_{0}+2n\pi}{k-\omega/v_{0}}$ with $n\in\mathbb{Z}$.
\begin{figure}
\includegraphics[width=1\columnwidth]{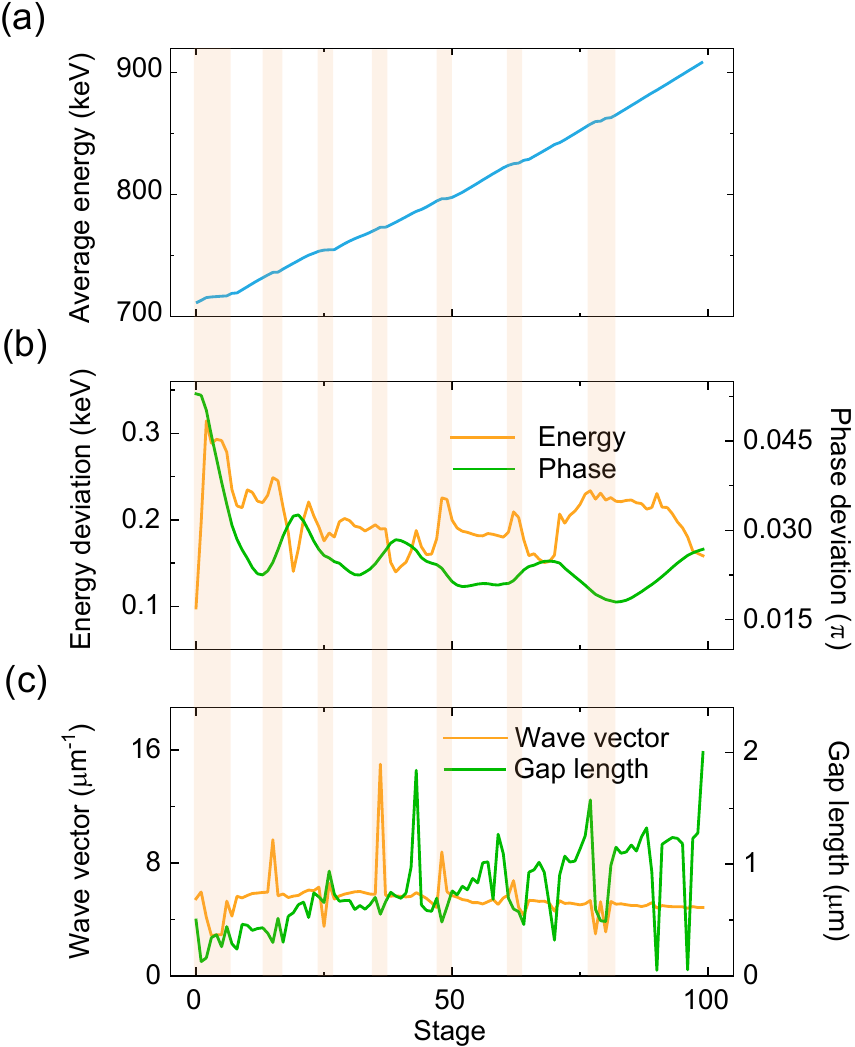}
\caption{The evolution of electrons in the multi-unit accelerator. (a), (b) are the average energy, the standard deviations of energy and phases for the electrons with  energy gain exceed $200\,\mathrm{keV}$. (c) the optimized parameters of each unit.
}
\label{Fig3}
\end{figure}
When the STM condition is satisfied, the particle would experience a constant and maximum force. Just as shown in the inset of Fig.\ref{Fig1}(d), the particle moving while the potential profile also varies with time, as shown by the arrows and also solid and dotted lines for different instants ($t_{1}$ and $t_{2}$). When $k=\omega/v$, as indicated by the yellow dot, the particle's position concerning the potential profile remains unchanged at different times. If $\phi_{0}=3\pi/2$ is satisfied initially, the particle will stay in the position where the gradient of the potential is maximum. Thus, the potential would always push the particle to accelerate it.

Although the proportional relation between the optimal achievable energy gain ($\delta E$) and $L$ indicates that a significant acceleration could be realized by a single modulation unit under the STM condition, the first-order approximation would be invalid for large $L$ when self-energy is significantly changed. So, the achievable energy gain in a single modulation unit is limited eventually. Therefore, it is necessary to go beyond the first-order approximation of $\delta E$ and optimize the $U\left(x,t\right)$ to achieve a global STM at every location for a given particle's initial parameters. This could be achieved by regulating the $A$, $k$ or $\phi_{0}$ to be spatially-temporally dependent. However, this continuously varying the field is experimentally challenging, we resorted to the approach that divides the potential into multiple units, as shown by Fig.~\ref{Fig1}(e), and thus achieve the STM by optimizing the parameters of each unit to make sure that the gradient of the field is always maximum to the particles. Just as it shows in Fig.~\ref{Fig1}(d), $A$ is chosen to be $0$ in the gray area to adjust its relative phase to the field before the particle interacts with the potential field in the next unit, then cascaded units would provide an accumulative effect. A similar idea was proposed and demonstrated in Ref.~\citep{Fan2019} to utilize a sequential of individually suspended waveguide to realize the accumulation of frequency shift.

Comparing the different trajectories of the inset in Fig.~\ref{Fig1}(d), for the same potential, only the particles with selected parameters, i.e. $v$ and phase $\phi_{0}$ when the particle enters into the potential, could be effectively accelerated, while the other particles might gain no energy or even be decelerated. This indicates an important property of the spatial-temporal accelerator: the acceleration is only valid to a portion of particles in an ensemble. In the experiment, the initial velocity or arriving time of particle ensembles usually has uncertainty that follows certain distributions, therefore, it is difficult to determine the parameters of fields that can achieve the best acceleration of the whole ensemble. In this work, we set the goal as accelerating or decelerating the particles to a certain threshold of velocity, while optimizing the probability of the particles in a given ensemble that could exceed the threshold. This problem may be solved by separately optimizing each unit with a greedy algorithm to obtain a globally
optimal solution.  In the following, we study the massive particle accelerator for electrons by the optical near-fields of photonic grating nanostructures. The idea is also generalized to the atom cooling and trapping through the AC Stark potential imposed by nano-structured optical fields, with more detail provided in Ref.~\cite{sm}.

Figure~\ref{Fig2}(a) sketches the integrated grating structures for electron acceleration. Each section of the grating provides the evanescent electromagnetic field when shining by a laser beam, which was experimentally studied in Ref.~\citep{Breuer2013,Peralta2013}. The oscillation frequency of the electric field $\omega$ is determined by the input laser, and the wavevector of the corresponding potential ($k$) is controllable by choosing an appropriate grating period. Hence, the STM could be realized when the electron and the spatial-temporal modulation field evolve at the same velocity as $v=\omega/k$. Meanwhile, the phase $\phi_{\mathrm{0}}$ could be directly controlled by an optical delay. Therefore, the photonic chip provides a flexible platform to integrate hundreds of units that satisfy the STM condition for considerable electron acceleration on $\mathrm{cm}$-scale chip {[}Fig.~\ref{Fig2}(a){]}.

According to Eq.~(\ref{worldline}), the kinetic energy change of electron in each unit is $\delta E=-e\mathcal{E}v\int_{t_{0}}^{t_{0}+\delta t}\sin(\omega t-kvt+\phi_{\mathrm{0}})dt$ to the first-order approximation, with $\mathcal{E}$ is the amplitude of the induced evanescent field. To determine the parameters of each unit for STM, the recurrence relation of energy and phase distribution of an electron ensemble between two adjacent units could be derived analytically. For example, the result of Ref.~\citep{Peralta2013} for a single unit could be derived accordingly~\citep{sm}. However, the analytical recurrence relation for cascaded units becomes highly nonlinear and it is impractical to derive the optimal parameters. Thus, in the following studies, the Monte Carlo method, as well as the greedy algorithm, are employed to optimize the parameters of 100 units.

In a simplified design, the length of each unit is set to be identical while the gap length between adjacent units is changed to control the relative arriving time (i.e. the phase) of electrons entering each unit. Meanwhile, the wavevector of each unit is optimized for achieving the STM condition. We set a threshold of energy gain and the objective of the optimization is the number of electrons that could be accelerated above the threshold. The results are summarized in Fig.~\ref{Fig2}(b), for a given ensemble of $5\times10^4$ electrons with random initial phases and an initial average electron energy of $711\,\mathrm{keV}$ (including the mass-energy $mc^{2}$), and the standard deviation of the initial energy is $0.142\,\mathrm{keV}$. Figure~\ref{Fig2}(b) shows the proportion of the electrons that could reach the threshold as indicated by the horizontal axis, and we found that the proportion reduces with the threshold increasing. It is because the theoretical up-bound of energy gain by $100$ units is limited to $260\,\mathrm{keV}$ according to Eq.~(\ref{eq:STMC}). The results indicate that as high as $30\%$ of the input electrons could be accelerated to half of the up-bound. For typical thresholds of $150\,\mathrm{keV}$ and $200\,\mathrm{keV}$, which correspond to atom velocity acceleration from $0.695c$ to $0.805c$ and $0.827c$ with $c$ is the vacuum light velocity, details of the final energy distribution of output electrons are plotted in Fig.~\ref{Fig2}(c). The results show that with a device total length less than $6\,\mathrm{mm}$, the electrons could be accelerated to $0.805c$ and $0.827c$ with the probability of $26.7\%$ and $6.9\%$, respectively. We also found that although the acceleration is probabilistic for the ensemble, the accelerated electrons have a very narrow energy spectrum and thus be distinguishable from others [as shown by Red and Blue shadow in Fig.~\ref{Fig2}(c)], in sharp contrast to the broad output energy distribution by a single acceleration unit~\citep{Breuer2013,Peralta2013} or an array of periodically aligned acceleration units~\citep{Sapra2020,Shiloh2021a}. So the output electrons with the energy above the threshold could be almost deterministically selected by choosing an appropriate time window, and such a property of the device is potentially useful in the future.

To interpret the cooperation of these units for massive particle acceleration, we plot the detailed average energy and standard deviation of energy and phases at different units in Fig.~\ref{Fig3}, for the output electrons whose energy gain exceeds 200\,$\mathrm{keV}$. The general trend, that wavevector reduces and gap length increases with the number of stages, indicates the varying STM condition for the accelerated electrons as the electron velocity increases. We call them acceleration units, where the average energy of the electron ensemble continuously increases. However, the parameters also show sudden jumps in Fig.~\ref{Fig3}(c), and the average energy in these units usually stays unchanged or decrease. As shown by the orange shadow in Figs.~\ref{Fig3}(a)-(c),  jumps in wavevector always correspond to the inflection point of energy standard deviation [orange curve in Fig.~\ref{Fig3}(b)], and have an influence on phase standard deviation [green curve in Fig.~\ref{Fig3}(b)]. We can conjecture that these jumps are aimed to compress the electron ensemble for narrower energy and phase distributions, thus these units are named as focus units. In the whole process, acceleration units cause continuously energy gain and divergence of the electrons. Thus several focus units are introduced to compress the electrons and make most of them satisfy STM conditions.

\begin{figure}
\includegraphics[width=1\columnwidth]{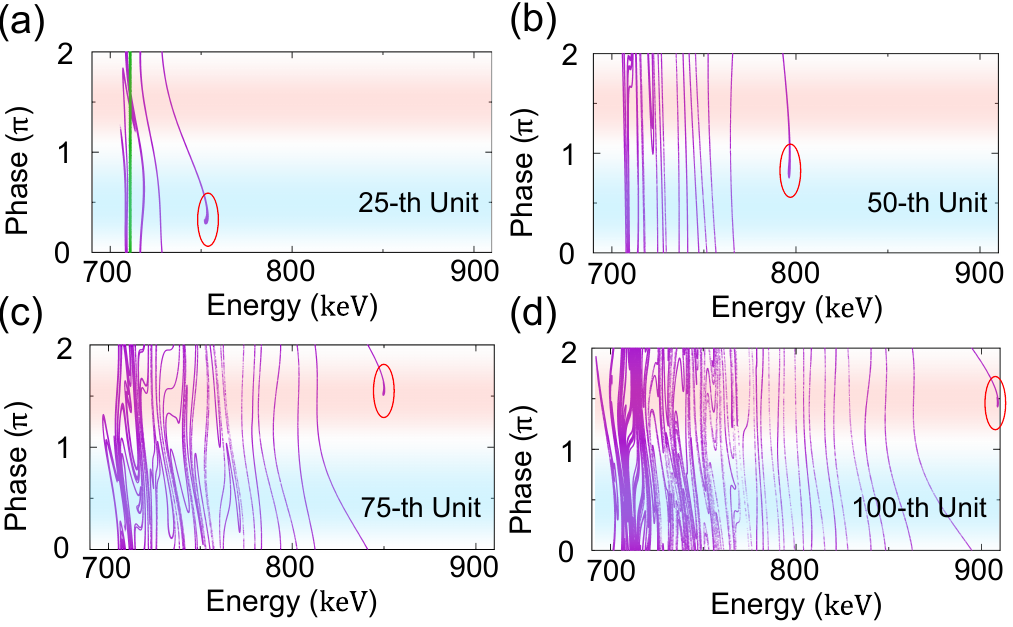}
\caption{The energy and phase distribution of electron ensemble entering 25-th (a), 50-th (b), 75-th (c), and 100-th (d) units. Green dots in (a) shows initial distribution of electrons. The parameters of each units are optimized for energy gain of $200\,\mathrm{keV}$, and other parameters are the same as those in Fig.~\ref{Fig2}(b). The red and blue areas correspond to the parameter regimes where electrons are accelerated and decelerated, respectively.
}
\label{Fig4}
\end{figure}

Furthermore, the states of the ensemble that enter certain units are depicted by the Poincare surface of section (SOS) to provide an intuitively interpret the electron evolution in two types of units. As presented in Fig.~\ref{Fig4} are the distributions of the energy and the phase of each electron in the ensemble for 25th, 50th, 75th and 100th units. The nonlinear mapping of the electrons state gives rise to the chaotic manifolds in the phase space, and a bunch of dots at the rightmost (in the red circle) corresponding to the portion of electrons could be effectively accelerated, while the remaining electrons are randomly distributed concerning phase. In the SOS, as indicated by the background color in Fig.~\ref{Fig4}, electrons could be accelerated (in the red region) while decelerated (in the blue region) depending on their phase when entering the unit. In focus units, such as 25-th and 50-th units [Figs.~\ref{Fig4}(a),(b)], the small ensemble of electrons located at the blue area to focus itself and in acceleration units, such as 75-th and 100-th units [Figs.~\ref{Fig4}(c),(d)], these electrons located at the center of the red area to gain maximum energy.

In conclusion, by generalizing the spatial-temporal modulation of optical photons to massive particles, we propose a very compact and efficient electron accelerator. On the photonic chip, electrons could be accelerated by $200\,\mathrm{keV}$ with the probability of $20\%$ at the range less than $6\,\mathrm{mm}$. The matched spatial-temporal modulation principle is universal and could be applied to other massive particles, for example, we proposed an integrated atom slower by a series of on-chip optical dipole trap tweezers to reduce atomic kinetic energy~\citep{sm}. For atom beams with initial velocity centered at $10\,\mathrm{m/s}$ can be trapped with probability of $20\%$ at the length less than 2mm.  Benefiting from the developing integrated photonic technologies, the distributed spatial-temporal modulation could be realized to efficiently manipulate particles on the chip and promises novel functional photonic devices.

\smallskip{}

\begin{acknowledgments}
This work was funded by the National Key Research and Development Program (Grant No.~2017YFA0304504) and the National Natural Science Foundation of China (Grant Nos.~11874342, and 11922411, 12104441, and U21A6006), Anhui Provincial Natural Science Foundation (Grant No.~2108085MA22), and also the Fundamental Research Funds for the Central Universities. The numerical calculations in this paper were partially done on the supercomputing system in the Supercomputing Center of University of Science and Technology of China.
\end{acknowledgments}

\clearpage{}

\twocolumngrid
\renewcommand{\thefigure}{S\arabic{figure}}
\setcounter{figure}{0} 
\renewcommand{\thepage}{S\arabic{page}}
\setcounter{page}{1} 
\renewcommand{\theequation}{S.\arabic{equation}}
\setcounter{equation}{0} 
\setcounter{section}{0}

\begin{center}
\textbf{\textsc{\LARGE{}Supplementary Material}}{\LARGE\par}
\par\end{center}

\tableofcontents{}

\section{Atom cooling}\label{Atom}

\begin{figure*}
\includegraphics[width=1.6\columnwidth]{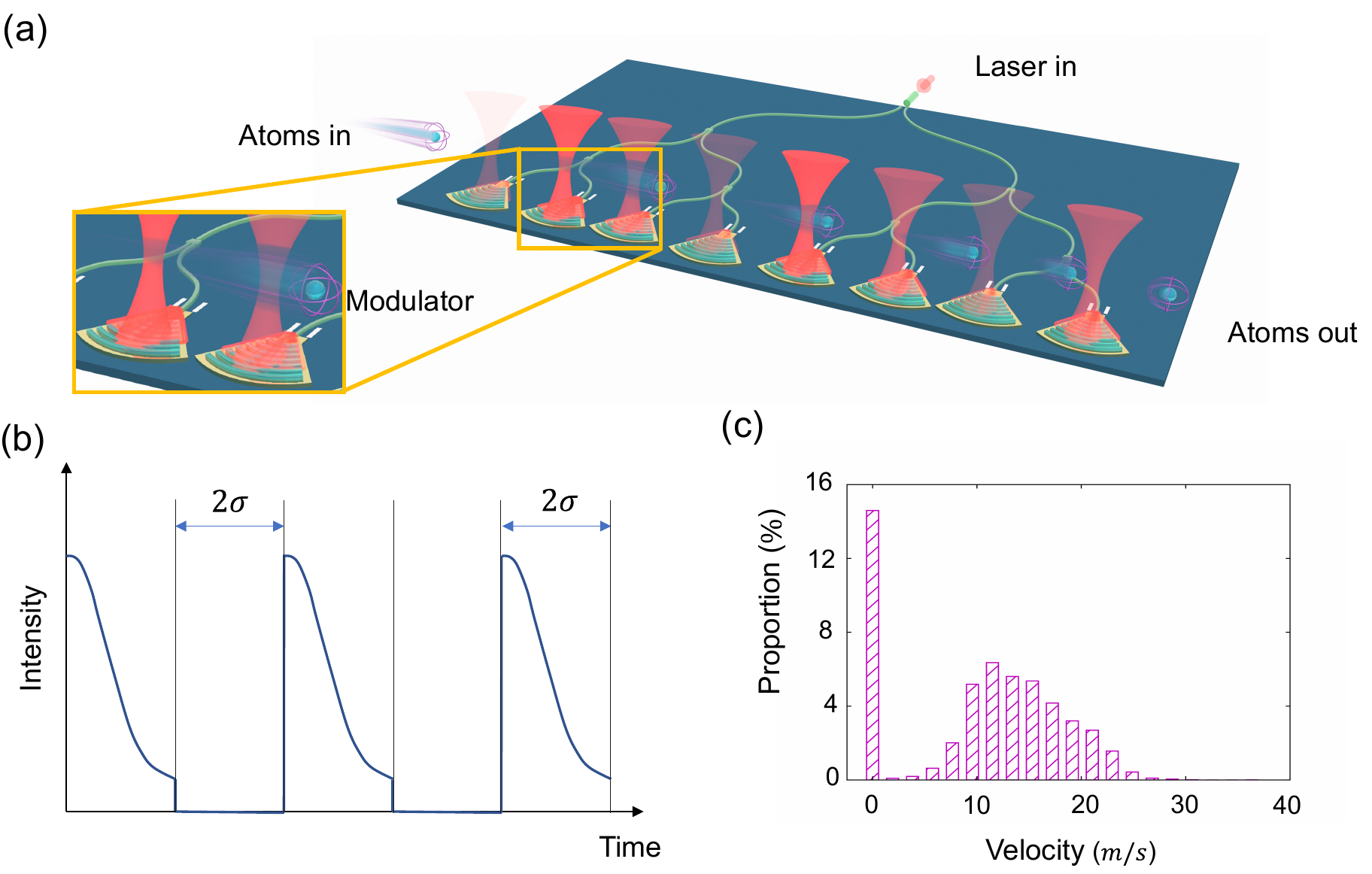}
\caption{(a) Sketch of the experimental setup for cascaded atom decelerator on a photonic chip. Atoms are emitted from left, and they interact with the spatial-temporally modulating dipole trap potential by the Gaussian beams. (b) The time-dependent intensity of input lasers to the gratings. The laser is modulated to be a series of pulse of half Gaussian function shape, with the standard deviation of the Gaussian function is $\sigma$. (c) Histogram statistic for velocity of an ensemble of atoms after deceleration. The central electric field is $2\times10^{6}\,\mathrm{V/m}$, the gap-length is set at $7.5\,\mathrm{\mu m}$, the laser detuning with respect to the atomic transition is 100\,nm. The total number of atoms in the ensemble is 10000, with their initial velocities being Gaussian distribution, which is centered at 10\,m/s with a standard deviation of 0.0625\,m/s.}
\label{figS1}
\end{figure*}

This spatial-temporal method for massive particle acceleration can also be adopted for atoms cooling by modulation of the spatial-temporary dependent optical dipole potential on a photonic chip. For example, a Gaussian beam of $880\,$nm laser, with its wavelength being $100\,$nm longer than the D2 transition of Rubidium atoms, could provide the attractive dipole potential for the Rubidium atoms. As schematically shown in Fig.~\ref{figS1}(a), integrated grating couplers could convert waveguide mode to free-space Gaussian beam (waist $W$) above the chip, with the laser amplitude in the waveguide are periodically modulated as $\mathcal{E}(t)=0$ for $-2\sigma\leq t<0$ and $\mathcal{E}(t)=\exp[-t^{2}/\sigma^{2}]$ for $0\leq t<2\sigma$, with the period of $4\sigma$ (See Fig.~\ref{figS1}(b)). For the laser frequency largely detuned from the atomic transition, the beams generate trap potential for atoms via AC-Stack effect that shifts the ground state energy as
\begin{equation}
    U=\frac{2\hbar\Delta\Omega_{R}^{2}\left(t\right)}{4\Delta^{2}+\Gamma^{2}+\Omega_{R}^{2}\left(t\right)},
\end{equation}
with $\Omega_{R}\left(t\right)=d\mathcal{E}\left(t\right)/\hbar$, $d$ is the dipole moment for the atomic transition. Therefore, the Gaussian beam can change the atom kinetic energy as
\begin{equation}
   \delta E\left(t_{0},t_{1}\right)=-\int_{t_{0}}^{t_{1}}\frac{4\hbar\Omega_{\mathrm{R}}^{2}(t)\Delta v\left|x_{0}+vt\right|}{\left[4\Delta^{2}+\Gamma^{2}+\Omega_{\mathrm{R}}^{2}(t)\right]W^{2}}dt,
\end{equation}
with $v$ is the velocity of the atom, $x_{0}$ corresponds to the atom position at $t_{0}$ with respect to the center of the Gaussian beam.

Considering an pre-cooled $^{87}\mathrm{Rb}$ atom ensembles with initial average velocity $10\,\mathrm{m/s}$ and standard deviation $0.0625\,\mathrm{m/s}$, we could use the chip depicted in Fig.~\ref{figS1}(a) for on-chip atom cooling and trapping. By the modulation scheme shown in Fig.~\ref{figS1}(b), the STM could be realized by optimizing the switching time $T$ that could be controlled by on-chip delay and the $\sigma$, with the two parameters corresponding to the phase and wavevector in Fig.~2 of the main text. The other parameters are fixed, with the laser waist, maximum potential depth and the distance between two adjacent gratings are $1.5\,\mathrm{\mu m}$, $3.5\,\mathrm{mK}$, and $7.5\,\mathrm{\mu m}$, respectively. Similar to the electron accelerator, we assume a device compose of 200 units, with $T$ and $\sigma$ can be optimized for each stage. The results of atom trapping are summarized in Fig.~\ref{figS1}(c). By such a $\sim2$\,mm-long photonic device on a chip, atoms can be cooled down and trapped with a probabilistic of about $27\%$ at the last ten stages. It is noticed that in atom cooling, we also observed the energy symmetry broken.

\section{Photon frequency shift}\label{Photon}

Under the spatial-temporal modulation, the  refractive index of a waveguide is inhomogenous, and the head of a wave may travel slower (faster) than its tail, which will induces a
frequency shift~\citep{Fan2016}. Choose two points in the wave which separate a wavelength, the shift of wavelength is
\begin{equation}
d\lambda=\Delta v dt=-\frac{c}{n_{\mathrm{eff}}^{2}}\frac{\partial \Delta n_{\mathrm{eff}}(x,t)}{\partial x}\lambda\frac{n_{\mathrm{eff}}}{c}dx
\end{equation}
where $n_{\mathrm{eff}}$ is the effective refractive index of the waveguide and $\Delta n_{\mathrm{eff}}(x,t)$ is the mechanically-induced spatial-temporal
modulation of $n_{\mathrm{eff}}$, $c$ is the speed of light in vacuum, $\lambda$ is the wavelength of photon, which is assumed to be a constant in a single unit. Specifically,
\begin{equation}
d\lambda=-v_{g}\frac{2\pi}{\omega}\frac{\partial \Delta n_{\mathrm{eff}}(x,t)}{\partial x}d x
\end{equation}
here $v_{g}$ is the light group velocity in the waveguide, $\omega$ is the light angular frequency. Then the frequency shift
\begin{equation}
d\omega=v_{g}\frac{2\pi}{\lambda}\frac{\partial \Delta n_{\mathrm{eff}}(x,t)}{\partial x}dx=v_{g}k_{0}\frac{\partial \Delta n_{\mathrm{eff}}(x,t)}{\partial x}dx
\end{equation}
where $k_{0}$ is the vacuum photon wavevector.
As the frequency shift is
adiabatic, which means that the amplitude change of the electromagnetic
wave is very slow, the photon number is conserved~\citep{Notomi2010}, thus $dE=\hbar d\omega$, with $\hbar$ is the reduced Plank constant. Total energy change in a single unit
\begin{equation}
\delta E\approx\hbar v_{\mathrm{g}}k_{0}\int d\zeta\frac{\partial}{\partial x}\left[\Delta n_{\mathrm{eff}}\left(x,t\right)\right].\label{eq:3}
\end{equation}
here, the integration is along the trajectory $\zeta$ from the initial wave packet position $\{x_0,t_0\}$ to finial wave packet position $\{x_1,t_1\}$, as in the main text.

\section{Analytical distribution evolution of electrons }\label{Analytical distribution evolution of electrons}

For an input electron ensemble, with $f_{n}(E_{n},\phi_{n})$ is
the distribution of energy and phase of electron that entering $(n+1)$-th
unit, where $E_{n}$ and $\phi_{n}$ is the corresponding energy and
phase, respectively.The mapping between inputs at $\left(n+1\right)$-th
and $\left(n+2\right)$-th unit is described by $E_{n+1}=E_{n}+\delta E,$
$\text{\ensuremath{\phi_{n+1}=\phi_{n}+\frac{L}{v}\omega+\Delta\phi_{n,n+1}}}$,
where $\Delta\phi_{n,n+1}$ is the controllable phase difference.
Therefore, the evolution of the distribution follows
\begin{equation}
f_{n+1}(E_{n+1},\phi_{n+1})=|\text{\ensuremath{\frac{\partial(E_{n},\phi_{n})}{\partial(E_{n+1},\phi_{n+1})}}}|f_{n}(E_{n},\phi_{n}).
\end{equation}
However, sometimes we are more interested in the distribution of energy
gain during the $n$-th accelerator, which we will denote as $g_{n}(\Delta E_{n})$,
here $\Delta E_{n}$ is defined relatively to the mean energy of the
electron ensemble
\begin{equation}
\overline{E}=\int E_{n-1}f_{n-1}\left(E_{n-1},\phi_{n-1}\right)d\phi_{n-1}dE_{n-1}.
\end{equation}
So, $g_{n}$ can be calculated through $f_{n-1}$ by
\begin{equation}
g_{n}(\Delta E_{n})=\int_{\mathrm{contour}}ds\frac{f_{n-1}(E_{n-1},\phi_{n-1})}{|\nabla(\Delta E_{n})|}.\label{eq:4}
\end{equation}
The integral is calculated along the contour with respect to $\Delta E_{n}(E_{n-1},\phi_{n-1})$
in the $\{E_{n-1},\phi_{n-1}\}$ phase space. For the first accelerator
unit, the time electrons arriving at the accelerator is usually uncontrollable
compared to the period of the oscillated field, therefore, the we
assume the phase has a independent, uniform distribution, i.e. $f_{0}(E_{0},\phi_{0})=\frac{1}{2\pi}\rho(E_{0})$.
In this case, we obtain
\begin{align}
g_{1}(\Delta E_{1}) & =\frac{1}{2\pi}\int\rho(E_{0})dE_{0}\frac{1}{\left|\partial\Delta E_{1}/\partial\phi_{0}\right|}.\label{eq:5}
\end{align}
according to Eq.~(\ref{eq:4}), with $v$ and $E_{\mathrm{0}}$ satisfies
relativistic relation $v=c\sqrt{1-m^{2}c^{4}/E_{0}^{2}}$. When we
take $\rho(E_{0})$ as a Gaussian distribution, Eq.~(\ref{eq:5})
agrees to the result of \citep{Peralta2013} and Monte Carlo simulation.

\end{document}